\documentclass[11pt,a4paper]{article}

\usepackage[utf8]{inputenc}
\usepackage[T1]{fontenc}
\usepackage[english]{babel}
\usepackage{amsmath,amssymb,amsthm}
\usepackage{graphicx}
\usepackage{booktabs}
\usepackage{multirow}
\usepackage[margin=2.5cm]{geometry}
\usepackage[colorlinks=true,linkcolor=blue!50!black,citecolor=blue!50!black,urlcolor=blue!50!black]{hyperref}
\usepackage{algorithm}
\usepackage{algpseudocode}
\usepackage{xcolor}
\usepackage{caption}
\usepackage{subcaption}
\usepackage{listings}
\usepackage[expansion=false]{microtype}
\usepackage{enumitem}
\usepackage{authblk}
\usepackage[round,authoryear]{natbib}

\newtheorem{definition}{Definition}
\newtheorem{property}{Property}
\newtheorem{example}{Example}

\title{\bf GPU-Accelerated Construction of the Gabriel Graph:\\
Parallelization Strategies and Applications to\\
Large-Margin Classification\\
and Graph-Based Clustering}

\author[1]{Diego Sanches Nere dos Santos}
\author[2]{Luiz Carlos Bambirra Torres}
\affil[1]{Department of Computing and Systems, Institute of Exact and Applied Sciences, Federal University of Ouro Preto (UFOP), João Monlevade, MG, Brazil}
\affil[2]{Department of Computing, Institute of Exact and Biological Sciences, Federal University of Ouro Preto (UFOP), Ouro Preto, MG, Brazil}

\date{}

\begin{document}
\maketitle

\begin{abstract}
The Gabriel graph is a proximity graph widely used in computational geometry and geometric machine learning, underlying large-margin classifiers and, as we show in this work, density-aware clustering methods. Its construction, however, has cubic time complexity $O(N^3)$ in the naive formulation, which severely limits its use on datasets beyond a few thousand samples. This paper presents a set of GPU parallelization strategies, implemented in CUDA, that make Gabriel graph construction practical at scale. We introduce (i) an implicit closed-form mapping from a linear thread index to the corresponding vertex pair $(i,j)$, which removes the need for an explicit $O(N^2)$ adjacency structure and lets every thread locate its pair in constant time and constant memory; (ii) a two-pass, bitmask-based, atomically-compacted edge-list construction that produces a dense edge array proportional only to the number of edges actually found; and (iii) a thread-coarsening strategy in which each thread evaluates several candidate pairs, trading launch and scheduling overhead against load balance. We evaluate these strategies on the acceleration of a Gabriel-graph-based large-margin classifier (NN-clas) and on a novel Gabriel-graph clustering algorithm that identifies connected components on a residual graph and performs GPU-parallel Union--Find with hierarchical edge-based merging. On 14 public benchmark datasets, the GPU implementation reduces Gabriel graph construction time by two to nearly four orders of magnitude relative to a sequential CPU baseline (up to $8484\times$), without loss of classification accuracy, and the proposed clustering method is statistically competitive with DBSCAN on the topological validity index Dunn while eliminating the need for an explicit density radius parameter.
\end{abstract}


\section{Introduction}
\label{sec:intro}

Proximity graphs occupy a central role in geometric approaches to machine learning: they turn a set of feature vectors into a combinatorial structure whose edges encode a notion of ``closeness'' that can be exploited for classification, clustering, dimensionality reduction, or anomaly detection. Among proximity graphs, the \emph{Gabriel graph}, introduced by \citet{gabrielsokal1969} in the context of geographic variation analysis, is particularly attractive because it depends on nothing but pairwise Euclidean distances, requires no free parameters (unlike $k$-nearest-neighbor or $\varepsilon$-radius graphs), and is a subgraph of the Delaunay triangulation, which guarantees several useful connectivity and locality properties \citep{matulasokal1980}.

A vertex pair $(x_i, x_j)$ is connected by an edge in the Gabriel graph if and only if the open hypersphere having the segment $\overline{x_i x_j}$ as diameter contains no other point of the dataset. This purely geometric criterion has made the Gabriel graph a popular structural backbone for large-margin classifiers \citep{garcia2015,ariasgarcia2024,hanriottorresbraga2024}, where edges connecting samples of different classes -- the \emph{support edges} -- approximate the decision margin between classes without solving any optimization problem.

The price of this elegance is computational: verifying the empty-hypersphere condition for every candidate pair against every other point costs $O(N^3)$ in the number of samples $N$, since there are $O(N^2)$ candidate pairs and each check is itself $O(N)$. This cost quickly becomes the dominant bottleneck of any Gabriel-graph-based pipeline once $N$ grows into the thousands, which is precisely the regime in which real-time or large-scale applications -- medical diagnosis, network intrusion detection, industrial monitoring -- would most benefit from these classifiers.

Because the empty-hypersphere test for a pair $(i,j)$ is completely independent of the test for any other pair $(k,l)$, Gabriel graph construction is an \emph{embarrassingly parallel} problem, and is therefore a natural candidate for acceleration on Graphics Processing Units (GPUs), whose Single-Instruction-Multiple-Threads (SIMT) architecture is designed precisely to exploit this kind of massive, independent, data-parallel work \citep{owens2008,nickolls2008}.

\paragraph{Contributions.} This paper isolates and details the GPU parallelization of Gabriel graph construction -- the computational core shared by every Gabriel-graph-based algorithm -- and studies it as a problem in its own right. Concretely, we contribute:

\begin{enumerate}[leftmargin=*]
    \item A closed-form, $O(1)$-time and $O(1)$-memory mapping from a one-dimensional CUDA thread index $k$ to the pair of vertex indices $(i,j)$ that it must evaluate, which avoids materializing an $O(N^2)$ index table (Section~\ref{sec:mapping});
    \item A two-pass, bitmask-based edge construction scheme that uses atomic operations to compact the sparse set of valid Gabriel edges into a dense array proportional to the true edge count, keeping GPU memory usage close to $O(|E|)$ rather than $O(N^2)$ (Section~\ref{sec:compaction});
    \item An empirical study of \emph{thread coarsening} -- assigning multiple candidate pairs to each thread -- showing that the best coarsening factor is dataset-dependent: it reduces execution time on one dataset but not on another (Section~\ref{sec:coarsening});
    \item Two downstream applications that validate the approach: an accelerated re-implementation of the NN-clas large-margin classifier \citep{ariasgarcia2024}, and a new GPU-native clustering algorithm that uses connected components of a distance-filtered Gabriel graph, with an automatic elbow-based cut and a GPU parallel Union--Find with hierarchical merging (Section~\ref{sec:applications});
    \item An experimental evaluation on 14 public datasets (synthetic and real), reporting speedups of up to $8484\times$ over a sequential CPU baseline and a statistical comparison of the resulting clustering quality against $K$-means and DBSCAN (Section~\ref{sec:results}).
\end{enumerate}

The remainder of the paper is organized as follows. Section~\ref{sec:related} reviews related work. Section~\ref{sec:background} introduces the Gabriel graph and the relevant CUDA/SIMT background. Section~\ref{sec:parallel} presents the parallelization strategies in detail. Section~\ref{sec:applications} describes the two applications used to validate the approach. Section~\ref{sec:results} reports the experimental results, Section~\ref{sec:discussion} discusses limitations, and Section~\ref{sec:conclusion} concludes with directions for future work.

\section{Related Work}
\label{sec:related}

\paragraph{Gabriel-graph-based classification.} \citet{gabrielsokal1969} originally proposed the Gabriel graph for geographic variation analysis; \citet{matulasokal1980} later characterized its geometric properties. In machine learning, the graph has been used to build hyperparameter-free, geometry-driven large-margin classifiers: edges connecting samples of opposite classes (support edges) delineate the separating margin without an explicit optimization procedure. \citet{garcia2015} studied the effect of label noise on such classifiers and proposed a quality-based filtering step, which we reuse as a pre-processing stage. \citet{ariasgarcia2024} proposed a hardware (FPGA) implementation of a Gabriel-graph classifier to reduce its computational cost, confirming that acceleration of this family of methods is an open and relevant problem, while \citet{hanriottorresbraga2024} extended Gabriel-graph classifiers to the multiclass setting and connected them to neural architectures, and \citet{ggclassification} provide a reference software implementation. None of these works targets general-purpose GPU acceleration of the underlying graph construction, which is the focus of the present paper.

\paragraph{GPU-accelerated machine learning.} The suitability of GPUs for data-parallel machine learning workloads is well documented \citep{owens2008,eon2021,zhang2023,colhak2025}. \citet{colhak2025} benchmark GPU versus CPU execution of machine learning libraries and report substantial throughput gains; \citet{zhang2023} present a GPU framework that couples neural simulation and AI at scale. These works motivate GPU acceleration in general but do not address the specific combinatorial structure of proximity-graph construction.

\paragraph{Thread granularity and coarsening.} \citet{magni2014} show that automatically tuning the amount of work assigned to each GPU thread -- \emph{thread coarsening} -- is essential, because a naive one-thread-per-element mapping can be dominated by scheduling overhead. We adopt this principle for Gabriel graph construction and quantify its effect empirically in Section~\ref{sec:results}.

\paragraph{Graph-based clustering.} Clustering by connected components of a proximity graph is a classical idea \citep{cormen2002,jain2010}; the difficulty lies in choosing which edges to keep. We adopt an elbow-based automatic threshold \citep{thorndike1953} on the sorted Gabriel edge weights, following the general taxonomy of internal cluster validity indices surveyed by \citet{ezugwu2022}, which we use throughout for quantitative evaluation.

\section{The Gabriel Graph}
\label{sec:background}

\subsection{Definition and Basic Properties}

Let $S = \{x_i\}_{i=1}^N \subset \mathbb{R}^n$ be a finite point set. The Gabriel graph of $S$ is the undirected graph $G=(V,E)$ with $V = \{1,\dots,N\}$ (one vertex per sample) and

\begin{definition}[Gabriel edge]
An edge $(i,j) \in E$ exists if and only if no other point $x_k \in S$, $k \neq i,j$, lies strictly inside the open hypersphere having $\overline{x_i x_j}$ as diameter. Equivalently,
\begin{equation}
\label{eq:gabriel}
\|x_i - x_j\|^2 \;\le\; \|x_i - x_k\|^2 + \|x_j - x_k\|^2, \qquad \forall\, x_k \in S \setminus \{x_i, x_j\}.
\end{equation}
\end{definition}

\begin{figure}[t]
    \centering
    \includegraphics[width=0.62\linewidth]{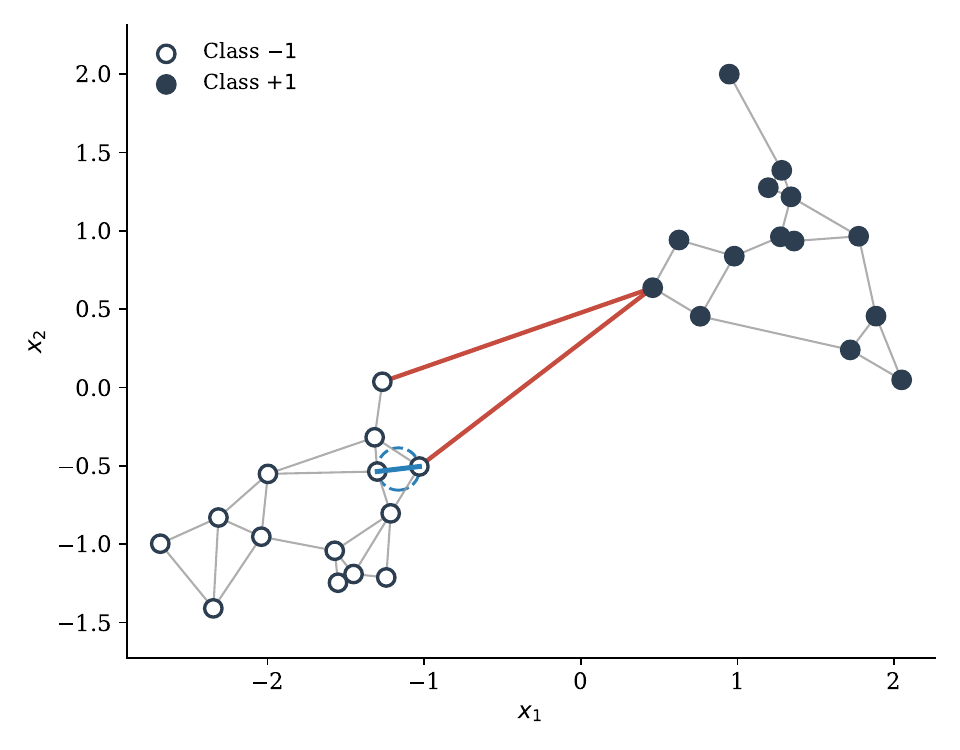}
    \caption{A Gabriel graph built on a synthetic two-class point set. Grey edges connect same-class neighbors; red edges are \emph{support edges} joining points of different classes. The dashed circle illustrates the empty-hypersphere test of Eq.~\eqref{eq:gabriel} for one admitted edge: no other point falls inside it.}
    \label{fig:gabriel_example}
\end{figure}

Figure~\ref{fig:gabriel_example} illustrates the construction: for the highlighted pair, the dashed circle (the two-dimensional analogue of the empty hypersphere) contains no other data point, so the edge is admitted. In a labeled binary classification setting $S = \{(x_i, y_i)\}$ with $y_i \in \{-1,+1\}$, we call \emph{support edges} the subset $E_S \subseteq E$ of edges whose endpoints have different labels; under mild assumptions on class overlap, $E_S$ localizes the decision boundary between classes \citep{gabrielsokal1969,ariasgarcia2024}.

\begin{property}[Complexity]
Checking Eq.~\eqref{eq:gabriel} for a single candidate pair costs $O(N)$ (one pass over all other points). Since there are $\binom{N}{2} = O(N^2)$ candidate pairs, the naive construction costs $O(N^3)$ in time. The graph itself is sparse: it is known to be a subgraph of the Delaunay triangulation, so $|E| = O(N)$ in the planar case and generally $|E| \ll N^2$ in higher dimensions for non-adversarial data.
\end{property}

The gap between the $O(N^3)$ construction cost and the $O(N)$--$O(N^2)$ size of the resulting graph is precisely what motivates a parallel, memory-conscious construction strategy: work is heavily redundant across candidate pairs (each requiring an independent scan of $N-2$ points) but the eventual output is small and sparse.

\subsection{The SIMT Execution Model}

CUDA organizes computation into a grid of thread blocks; each block contains threads grouped into \emph{warps} of 32 threads that execute the same instruction stream in lock-step (Single Instruction, Multiple Threads, SIMT) \citep{nvidia2025guide}. Performance is maximized when threads within a warp follow the same control-flow path (avoiding \emph{warp divergence}) and when global memory accesses are coalesced. Concurrent writes to shared memory locations are protected with atomic operations (e.g., \texttt{atomicAdd}, \texttt{atomicOr}, \texttt{atomicMin}), which guarantee indivisible read-modify-write semantics at the cost of potential serialization when many threads contend for the same address \citep{nvidia2025guide}.

Gabriel graph construction maps naturally onto this model: each candidate pair $(i,j)$ can, in principle, be assigned to an independent thread, with no data dependency between threads other than read-only access to the (immutable) point set. The remainder of this paper addresses \emph{how} to organize that mapping efficiently.

\section{Parallel Construction Strategies}
\label{sec:parallel}
\label{sec:mapping}

\subsection{Naive Parallelization and the Indexing Problem}

A direct GPU translation of Gabriel graph construction launches one thread per candidate pair $(i,j)$, $i<j$, and has each thread evaluate Eq.~\eqref{eq:gabriel}. The number of candidate pairs is
\begin{equation}
\label{eq:totalpairs}
T = \binom{N}{2} = \frac{N(N-1)}{2}.
\end{equation}
The immediate difficulty is indexing: CUDA launches a one-dimensional (or low-dimensional) grid of threads, each receiving a \emph{linear} global index $k \in [0, T-1]$; the thread must recover the pair $(i,j)$ that $k$ represents. Materializing a lookup table of size $T = O(N^2)$ defeats the purpose of an efficient implementation, since it alone can exceed available GPU global memory for large $N$ (e.g., $N=30{,}000$ already gives $T \approx 4.5\times 10^8$ pairs).

\subsection{Implicit Index Mapping \texorpdfstring{$k \mapsto (i,j)$}{k to (i,j)}}

We derive a closed-form inverse of the standard row-major enumeration of the strict upper triangle of the $N\times N$ pair matrix,
\[
(0,1),(0,2),\dots,(0,N-1),\;(1,2),(1,3),\dots,(1,N-1),\;\dots,\;(N-2,N-1),
\]
so that every thread can compute its own $(i,j)$ in constant time and constant memory, using only $N$ and its own index $k$.

Row $i$ contains $L_i = N-i-1$ valid pairs. The number of pairs enumerated strictly before row $i$ is
\begin{equation}
P(i) = \sum_{r=0}^{i-1} (N - r - 1) = \frac{N(N-1)}{2} - \frac{(N-i)(N-i-1)}{2}.
\end{equation}
Index $k$ belongs to the (unique) row $i$ satisfying $P(i) \le k < P(i+1)$. Solving this quadratic inequality for $i$ in closed form gives
\begin{equation}
\label{eq:i-formula}
i = N - 2 - \left\lfloor \frac{\sqrt{-8k + 4N(N-1) - 7}}{2} - 0.5 \right\rfloor .
\end{equation}
The column index then follows directly from the offset within row $i$:
\begin{equation}
\label{eq:j-formula}
j = k - \left(\frac{N(N-1)}{2} - \frac{(N-i)(N-i-1)}{2}\right) + i + 1.
\end{equation}
Equations~\eqref{eq:i-formula}--\eqref{eq:j-formula} let every thread invert its linear index $k$ into a vertex pair $(i,j)$ using a handful of arithmetic operations and one square root -- no auxiliary memory, and no per-thread branching that would introduce warp divergence.

\begin{example}
For $N=5$, $T=10$ and the enumeration is
\[
\begin{array}{c|ccccc}
 & 0 & 1 & 2 & 3 & 4\\\hline
0 & - & (0,1)_0 & (0,2)_1 & (0,3)_2 & (0,4)_3\\
1 &  & - & (1,2)_4 & (1,3)_5 & (1,4)_6\\
2 &  &  & - & (2,3)_7 & (2,4)_8\\
3 &  &  &  & - & (3,4)_9\\
4 &  &  &  &  & -
\end{array}
\]
For $k=5$, Eq.~\eqref{eq:i-formula} gives $i = 3 - \lfloor \sqrt{33}/2 - 0.5\rfloor = 3 - \lfloor 2.37\rfloor = 1$, then $P(1)=4$ and Eq.~\eqref{eq:j-formula} gives $j = 1 + 1 + (5-4) = 3$, correctly recovering the pair $(1,3)$.
\end{example}

To guard against 64-bit overflow when $N$ is large, all index arithmetic (grid sizing, $T$, and the intermediate terms of Eqs.~\eqref{eq:i-formula}--\eqref{eq:j-formula}) is carried out using 64-bit integer/floating-point types.

\subsection{Two-Pass Construction with Bitmask Compaction}
\label{sec:compaction}

Even without materializing an index table, naively allocating one flag per candidate pair still costs $O(N^2)$ memory if a full byte (or worse, a 32-bit \texttt{int}) is used per flag. We instead use a two-pass, bitmask-compacted scheme:

\begin{enumerate}[leftmargin=*]
    \item \textbf{Pass 1 -- Marking.} Each thread evaluates Eq.~\eqref{eq:gabriel} for its pair $(i,j)$ (recovered via Eqs.~\eqref{eq:i-formula}--\eqref{eq:j-formula}). If the pair satisfies the Gabriel condition, the thread sets a single bit in a global bitmask using \texttt{atomicOr}, reducing flag storage by a factor of 32 relative to one \texttt{int} per pair, and increments a global counter with \texttt{atomicAdd} to determine the exact number of valid edges.
    \item \textbf{Intermediate allocation.} The host reads the counter and allocates exactly the memory required for the dense edge list (and, if needed, the associated edge weights).
    \item \textbf{Pass 2 -- Compaction.} Threads re-scan the bitmask; for every set bit, an atomic increment on a write-cursor determines that edge's position in the dense output array, which is then populated.
\end{enumerate}

This scheme keeps peak GPU memory proportional to the number of candidate pairs only during the (cheap, bit-packed) marking pass, and proportional to $|E|$ thereafter -- the true asymptotic footprint of the graph. All device vectors are managed through the Thrust library (\texttt{thrust::device\_vector}), and intermediate buffers are explicitly released (\texttt{shrink\_to\_fit}) as soon as they are no longer needed, which is important because subsequent stages (noise filtering, classification, or clustering) may need to allocate further structures on the same device.

\subsection{Thread Coarsening}
\label{sec:coarsening}

\begin{figure}[t]
    \centering
    \includegraphics[width=0.62\linewidth]{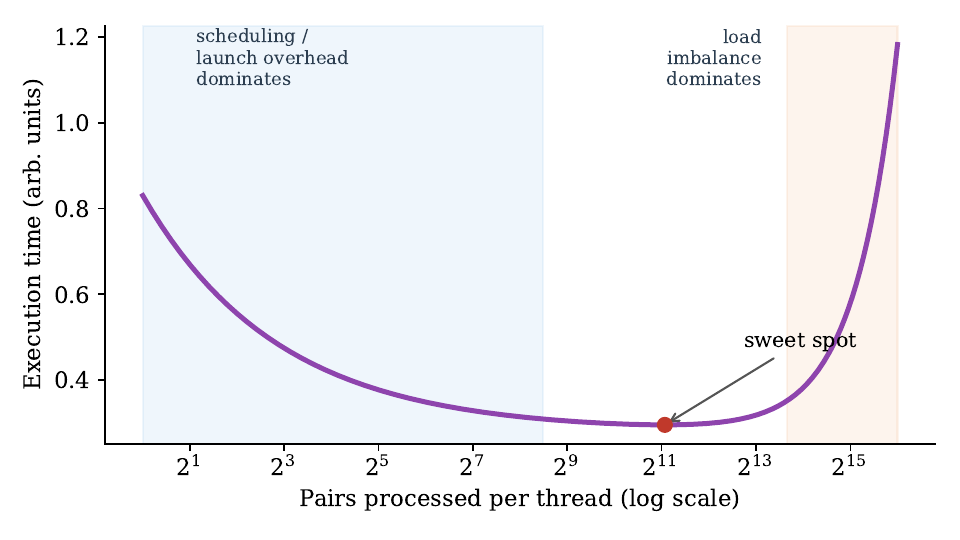}
    \caption{Schematic illustration of the thread-coarsening trade-off in Gabriel graph construction: very fine granularity is dominated by kernel-launch and scheduling overhead, while very coarse granularity is dominated by load imbalance across threads. An intermediate ``sweet spot'' minimizes total execution time; measurements on two datasets, where the optimum turns out to be dataset-dependent, are reported in Section~\ref{sec:coarsening-results}.}
    \label{fig:coarsening}
\end{figure}

Mapping exactly one candidate pair to one thread maximizes available parallelism but also maximizes the number of threads that must be created, scheduled and retired, which introduces overhead that is not always amortized by useful work -- especially for the many pairs, per warp, that fail the Gabriel test early. We therefore evaluate \emph{thread coarsening} \citep{magni2014}: each thread is assigned a contiguous block of $c$ candidate pairs (indices $k, k+1, \dots, k+c-1$ under the mapping of Eqs.~\eqref{eq:i-formula}--\eqref{eq:j-formula}), amortizing per-thread overhead and improving data locality, at the cost of a coarser and potentially less balanced distribution of work across the device. Figure~\ref{fig:coarsening} sketches the expected trade-off: execution time can fall as $c$ grows (overhead amortization dominates), reach a minimum, and rise again as load imbalance across threads dominates. Section~\ref{sec:coarsening-results} measures whether this holds in practice.

\subsection{Summary of the Construction Pipeline}

Algorithm~\ref{alg:pipeline} summarizes the full GPU construction pipeline combining implicit indexing, coarsening and two-pass compaction.

\begin{algorithm}[t]
\caption{GPU Gabriel graph construction with coarsening $c$}
\label{alg:pipeline}
\begin{algorithmic}[1]
\Require point set $\{x_i\}_{i=1}^N$ on device, coarsening factor $c$
\Ensure dense edge list $E$
\State $T \gets N(N-1)/2$ \Comment{Eq.~\eqref{eq:totalpairs}, 64-bit}
\State allocate bitmask $B$ of $\lceil T/32 \rceil$ words; zero-initialize; allocate counter $\gamma \gets 0$
\For{\textbf{each} thread $t$ \textbf{in parallel}} \Comment{Pass 1: marking}
    \For{$k = t\cdot c$ \textbf{to} $\min((t+1)c,T)-1$}
        \State $(i,j) \gets$ invert $k$ via Eqs.~\eqref{eq:i-formula}--\eqref{eq:j-formula}
        \If{Eq.~\eqref{eq:gabriel} holds for $(i,j)$}
            \State \texttt{atomicOr}$(B[k/32], 1 \ll (k \bmod 32))$
            \State \texttt{atomicAdd}$(\gamma, 1)$
        \EndIf
    \EndFor
\EndFor
\State allocate dense array $E$ of size $\gamma$ on host readback of $\gamma$; reset write-cursor $w \gets 0$
\For{\textbf{each} thread $t$ \textbf{in parallel}} \Comment{Pass 2: compaction}
    \For{$k = t\cdot c$ \textbf{to} $\min((t+1)c,T)-1$}
        \If{bit $k$ of $B$ is set}
            \State $p \gets$ \texttt{atomicAdd}$(w, 1)$
            \State $(i,j) \gets$ invert $k$; \quad $E[p] \gets (i,j)$
        \EndIf
    \EndFor
\EndFor
\State \Return $E$
\end{algorithmic}
\end{algorithm}

\section{Applications}
\label{sec:applications}

To validate the parallelization strategies of Section~\ref{sec:parallel}, we apply them to two Gabriel-graph-based algorithms: an existing large-margin classifier and a new clustering method.

\subsection{Accelerating a Gabriel-Graph Large-Margin Classifier}
\label{sec:nnclas}

NN-clas \citep{ariasgarcia2024} is a large-margin classifier that (i) builds the Gabriel graph of the labeled training set; (ii) applies a quality-based filtering step \citep{garcia2015} that removes noisy vertices whose neighborhood quality $q(x_i) = |\hat{G}r(x_i)|/|Gr(x_i)|$ falls below the class-wise average threshold; (iii) rebuilds the Gabriel graph on the filtered vertex set and extracts the support edges $E_S$; and (iv) classifies a new sample $x$ by a 1-nearest-neighbor rule restricted to the endpoints of $E_S$ (the ``border vertices''), rather than the full training set. Because both graph-construction passes (steps i and iii) dominate the algorithm's cost, they are the ones accelerated with the pipeline of Section~\ref{sec:parallel}; filtering and border-vertex classification remain comparatively cheap, embarrassingly-parallel, per-vertex operations executed on the GPU without further modification.

\subsection{A Gabriel-Graph Clustering Algorithm}
\label{sec:clustering}

Building on the intuition that short Gabriel edges reflect strong local connectivity while long edges tend to bridge distinct clusters, we propose a clustering algorithm that runs entirely on the GPU, avoiding host--device transfers once the graph has been built. It proceeds in four stages:

\begin{enumerate}[leftmargin=*]
    \item \textbf{Automatic threshold via the elbow heuristic.} Edge weights (Euclidean distances) are sorted (\texttt{thrust::sort}). Treating the sorted weights as a curve, a straight line is drawn between its first and last point, and a parallel kernel computes, for every point, its perpendicular distance to that line; the point of maximum distance is taken as the ``elbow'' and defines the cut threshold $T_{\text{corte}}$ \citep{thorndike1953}.
    \item \textbf{Residual graph.} Edges with weight $w \le T_{\text{corte}} + \varepsilon$ are kept via GPU stream compaction (\texttt{thrust::copy\_if} with zipped iterators); longer edges are dropped from the initial connectivity structure but retained in memory for the merging stage.
    \item \textbf{Parallel Union--Find.} Connected components of the residual graph are computed with a GPU parallel Union--Find: a \emph{hook} kernel processes residual edges in parallel and merges trees using \texttt{atomicMin} on the parent labels (the higher-ID root becomes a child of the lower-ID root, resolving write conflicts deterministically), followed by a \emph{path-compression} kernel that flattens every label to point directly at its root. The two kernels alternate until no label changes.
    \item \textbf{Hierarchical merge (optional).} If the number of connected components exceeds a user-specified target $K$, the algorithm repeatedly (a) finds, via a parallel reduction over \emph{all} original Gabriel edges (including those dropped in step 2), the minimum-weight edge that still connects two distinct components, and (b) relabels every point of the absorbed component to the surviving label, until exactly $K$ components remain or no inter-component edge exists.
\end{enumerate}

This design keeps the entire clustering pipeline -- from graph construction to final labels -- resident on the GPU, and requires no density radius or minimum-points parameter (unlike DBSCAN); when a target $K$ is not supplied, the algorithm reports the number of clusters implied by the elbow-selected threshold.

\section{Experimental Results}
\label{sec:results}

\subsection{Setup}

Two hardware configurations were used. For the sequential-versus-parallel classifier comparison: an AMD Ryzen 3 2200G CPU, an NVIDIA RTX 4070 Ti GPU (12 GB VRAM, 7680 CUDA cores), 16 GB RAM, Linux Mint 21.3, \texttt{nvcc} 11.5, G++-9, C++17. For the clustering validation, in order to test a more demanding hardware configuration: an AMD Ryzen 9 7900 (12 cores, 24 threads) CPU, an NVIDIA RTX 5070 Ti GPU (16 GB VRAM), 64 GB RAM, Ubuntu 24.04.3 LTS, \texttt{nvcc} 13.0. All timings isolate kernel and algorithmic execution time; host--device transfer overhead is not the focus of this study, since our goal is to characterize the relative acceleration and efficiency of the parallel kernels themselves.

\subsection{Sequential vs. Parallel Graph-Based Classification}

\begin{figure}[t]
    \centering
    \includegraphics[width=0.85\linewidth]{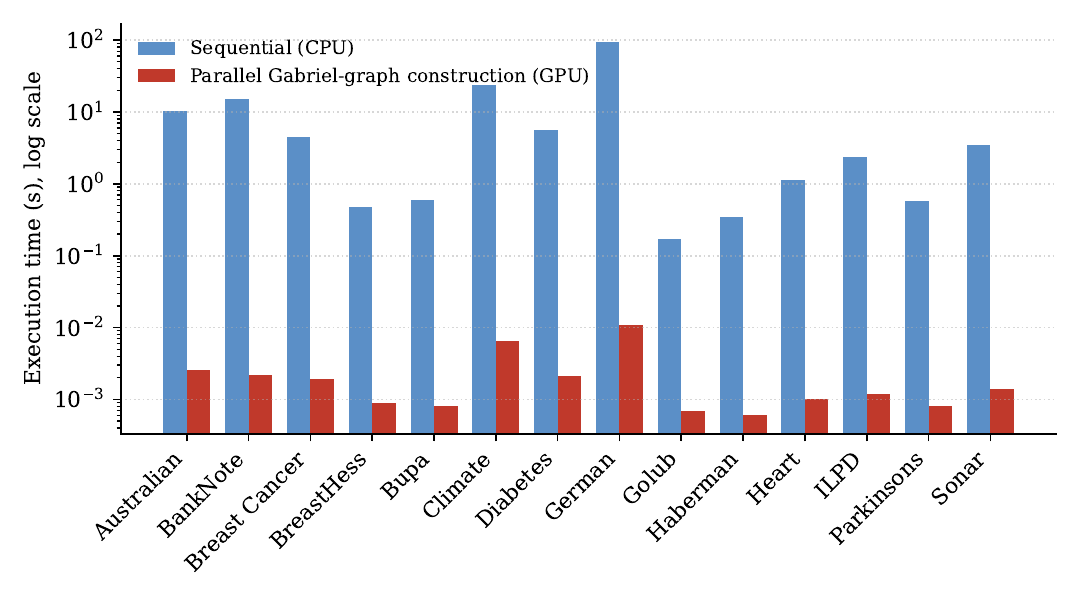}
    \caption{Execution time of Gabriel-graph-based classification, sequential (CPU) vs.\ GPU-parallel construction, log scale. Speedups range from $237\times$ (Golub, $N=72$) to $8484\times$ (German, $N=1000$).}
    \label{fig:speedup}
\end{figure}

Table~\ref{tab:speedup} reports execution time, speedup, accuracy, and F1-score for NN-clas on 14 public datasets (sizes 72--1372 samples, 3--60 features), comparing a sequential CPU baseline against the GPU pipeline of Section~\ref{sec:parallel} (one candidate pair per thread, i.e., coarsening factor $c=1$). Figure~\ref{fig:speedup} visualizes the same data on a logarithmic scale.

\begin{table}[t]
\centering
\caption{Execution time, speedup, accuracy and F1-score: sequential vs. GPU-parallel Gabriel-graph classifier.}
\label{tab:speedup}
\small
\begin{tabular}{lrrrrrrr}
\toprule
Dataset & $N$ & Feat. & Seq. (s) & Par. (s) & Speedup & Acc. & F1 \\
\midrule
Australian & 690 & 14 & 10.3789 & 0.0026 & $3933.8\times$ & 0.8487 & 0.8250 \\
BankNote & 1372 & 4 & 15.1423 & 0.0022 & $6882.5\times$ & 0.9978 & 0.9970 \\
Breast Cancer & 683 & 9 & 4.4336 & 0.0019 & $2333.8\times$ & 0.9648 & 0.9730 \\
BreastHess & 133 & 30 & 0.4752 & 0.0009 & $558.6\times$ & 0.7879 & 0.8510 \\
Bupa & 345 & 6 & 0.5933 & 0.0008 & $742.1\times$ & 0.5916 & 0.5360 \\
Climate & 540 & 18 & 23.4760 & 0.0064 & $3665.6\times$ & 0.8870 & 0.9380 \\
Diabetes & 768 & 8 & 5.5789 & 0.0021 & $2656.6\times$ & 0.7474 & 0.7980 \\
German & 1000 & 24 & 93.3308 & 0.0110 & $\mathbf{8484.6\times}$ & 0.7120 & 0.7780 \\
Golub & 72 & 50 & 0.1709 & 0.0007 & $237.4\times$ & 0.7732 & 0.8280 \\
Haberman & 306 & 3 & 0.3409 & 0.0006 & $532.6\times$ & 0.6431 & 0.7520 \\
Heart & 270 & 13 & 1.1135 & 0.0010 & $1080.1\times$ & 0.8222 & 0.8360 \\
ILPD & 579 & 10 & 2.3315 & 0.0012 & $1942.6\times$ & 0.6028 & 0.6960 \\
Parkinsons & 195 & 22 & 0.5699 & 0.0008 & $759.9\times$ & 0.8045 & 0.8740 \\
Sonar & 208 & 60 & 3.4717 & 0.0014 & $2548.3\times$ & 0.7355 & 0.6870 \\
\bottomrule
\end{tabular}
\end{table}

Speedup tends to grow with problem size, from $237\times$ on the smallest dataset (Golub, $N=72$, dominated by kernel-launch and host--device overhead) to $8484\times$ on German ($N=1000$, 24 features), the highest in the table. Accuracy and F1-score of the parallel implementation match the sequential baseline in every case (differences are within numerical noise), confirming that the atomic-based compaction and bitmask marking introduce no correctness loss. In absolute terms, execution time drops from the range of seconds -- in one case, over 90 seconds -- to low milliseconds, which is the difference between an offline-only method and one usable in near-real-time pipelines.

\subsection{Effect of Thread Coarsening}
\label{sec:coarsening-results}

We swept the coarsening factor $c$ (pairs per thread, in powers of two) on two larger public datasets: \emph{Diabetes} ($N\approx 100{,}000$ clinical records; $c \in \{1, 2, 4, \dots, 65536\}$) and \emph{Internet Firewall Data} ($N\approx 65{,}000$ network-traffic records, from the UCI Machine Learning Repository; $c \in \{1, 2, 4, \dots, 1024\}$). The two datasets behave differently.

On Diabetes, the trade-off sketched in Figure~\ref{fig:coarsening} appears clearly. Execution time decreases from $c=1$ to a minimum at roughly $128$--$512$ pairs per thread, about $15\%$ below the $c=1$ time, consistent with the amortization of per-thread creation and scheduling overhead over more comparisons. Beyond that optimum the time increases again, exceeds the $c=1$ time for $c \gtrsim 8192$, and is largest at $c=65536$, roughly $35\%$ above $c=1$. This is the pattern expected when coarser blocks are unevenly sized in useful work, so that some streaming multiprocessors finish early and idle while others are still processing their (larger, or Gabriel-condition-heavy) blocks.

On Internet Firewall Data, in contrast, execution time \emph{increases} with $c$ over the whole range tested: the best configuration is $c=1$, and $c=1024$ is roughly $40\%$ slower. The run-to-run variability is large relative to the differences between neighboring values of $c$, so only the overall trend should be read from this curve. We did not investigate the cause; plausible factors are the dimensionality of the data and the density of Gabriel edges (i.e., how often the early-exit condition in Eq.~\eqref{eq:gabriel} triggers), which change the per-pair cost and its variance across threads.

These measurements agree with the general thread-coarsening principle of \citet{magni2014} that the best granularity depends on the workload: for Gabriel graph construction, one pair per thread is not always optimal (Diabetes), but neither is a large per-thread batch, and coarsening can also bring no benefit at all (Firewall). We therefore recommend tuning $c$ empirically per dataset rather than fixing it a priori. The speedups in Table~\ref{tab:speedup} use $c=1$ and therefore do not include any gain from coarsening.

\subsection{Clustering Quality}
\label{sec:clustering-results}

We evaluated the Gabriel-graph clustering algorithm of Section~\ref{sec:clustering} against $K$-means and DBSCAN on 14 datasets (7 synthetic 2-D sets with non-convex structure -- e.g., \emph{Spiral}, \emph{Fullmoon}, \emph{Halfkernel} -- and 7 real UCI datasets with 214--30{,}000 samples and up to 57 features), using five internal validity indices: Silhouette, Davies--Bouldin, Calinski--Harabasz, Dunn, and Xie--Beni \citep{ezugwu2022}. Two regimes were tested: an \emph{automatic} regime, in which the elbow-selected threshold alone determines the number of clusters, and a \emph{$K$-optimized} regime, in which the hierarchical merge stage forces convergence to the ground-truth number of classes.

Across both regimes, a consistent pattern emerges. On variance-based indices that are biased toward convex, equal-radius clusters (Silhouette, Calinski--Harabasz), $K$-means is statistically superior, as expected given its explicit minimization of within-cluster variance. On the Dunn index, which rewards true topological separation and is sensitive to non-convex structure, the proposed method is statistically indistinguishable from DBSCAN (average rank $1.69$ vs.\ $1.73$ in the $K$-optimized regime, both significantly better than $K$-means under a Friedman test with Nemenyi and Bonferroni--Dunn post-hoc analysis at $\alpha=0.05$) while requiring no density-radius hyperparameter. The hierarchical merge stage materially improves the method's behavior on cohesion-oriented metrics: on Silhouette and Davies--Bouldin, the $K$-optimized variant becomes statistically indistinguishable from $K$-means (Bonferroni--Dunn critical difference $\mathrm{CD}=0.8472$), correcting the over-fragmentation observed under the purely automatic regime. These results indicate that a Gabriel-graph clustering method, made computationally tractable by the GPU construction pipeline of Section~\ref{sec:parallel}, is a viable density-aware alternative to DBSCAN for non-convex cluster shapes, at negligible additional hyperparameter cost.

\section{Discussion and Limitations}
\label{sec:discussion}

The results support two related claims. First, the $O(N^3)$ cost that has historically limited Gabriel-graph methods to small or medium datasets is not an intrinsic obstacle but an artifact of sequential execution: because the empty-hypersphere test is independent across candidate pairs, a GPU implementation combining implicit pair indexing and bitmask-compacted two-pass construction reduces execution time by two to nearly four orders of magnitude with no loss of downstream accuracy (Table~\ref{tab:speedup}, measured with $c=1$); thread coarsening can reduce it further on some datasets, but not on all. Second, once construction is fast, the Gabriel graph becomes a practical building block not only for classification (as in NN-clas) but also for density-aware clustering, where it is competitive with DBSCAN on the Dunn index while avoiding an explicit density-radius parameter.

Several limitations should be noted. The reported speedups compare a single-threaded CPU baseline to a single-GPU implementation on hardware with a substantial theoretical performance gap between host and device, which the community's convention (and our stated intent to isolate parallel-kernel efficiency) accepts but which should be kept in mind when transferring absolute speedup numbers to other hardware pairings; a multi-threaded CPU baseline (e.g., OpenMP) would narrow the reported gap while leaving the qualitative conclusions unchanged. The clustering method's hierarchical merge stage relies on a global minimum-weight-edge search per merge step, which, while GPU-parallel, adds sequential dependency across merge iterations and could itself be a target for further optimization (e.g., a parallel priority structure). Finally, the largest datasets used for the coarsening study ($N \approx 10^5$) already produce several billion candidate pairs ($T \approx 5\times 10^{9}$ for $N=10^5$); scaling further, into the range where the bitmask itself exceeds single-GPU memory, would require a multi-GPU or out-of-core extension of the pipeline described here.

\section{Conclusion}
\label{sec:conclusion}

We presented a set of GPU parallelization strategies for constructing the Gabriel graph -- implicit constant-time pair indexing, two-pass bitmask-compacted edge construction, and thread coarsening -- and showed that, applied to a Gabriel-graph large-margin classifier, they reduce execution time by up to $8484\times$ relative to a sequential CPU baseline across 14 public datasets, without any loss of classification accuracy. Because the same accelerated construction is reusable across tasks, we further proposed a new GPU-native clustering algorithm that filters the Gabriel graph by an automatically detected elbow threshold, extracts connected components via a parallel Union--Find, and optionally merges components hierarchically to reach a target cluster count; this method is statistically competitive with DBSCAN on the Dunn cluster-validity index while requiring no density-radius hyperparameter. Taken together, these results indicate that GPU acceleration removes the main practical obstacle -- cubic-time construction -- that has historically confined Gabriel-graph methods to small datasets, opening them up to time-sensitive, large-scale applications such as real-time medical diagnosis and network intrusion detection.

Future work includes multi-GPU and out-of-core extensions for datasets whose candidate-pair bitmask exceeds single-device memory, energy-efficiency characterization of the proposed kernels, alternative automatic cut-selection criteria beyond the elbow heuristic (e.g., statistically motivated thresholds), and a systematic comparison against multi-threaded CPU baselines to isolate the contribution of GPU parallelism from that of parallelism in general.

\section*{Acknowledgments}
This work is derived from an undergraduate thesis conducted in the Department of Computing and Systems, Federal University of Ouro Preto (UFOP), Brazil.

\bibliographystyle{plainnat}
\bibliography{references}

@article{gabrielsokal1969,
  author  = {Gabriel, K. R. and Sokal, R. R.},
  title   = {A New Statistical Approach to Geographic Variation Analysis},
  journal = {Systematic Zoology},
  volume  = {18},
  number  = {3},
  pages   = {259--278},
  year    = {1969},
  doi     = {10.2307/2412323}
}

@article{matulasokal1980,
  author  = {Matula, David W. and Sokal, Robert R.},
  title   = {Properties of {G}abriel Graphs Relevant to Geographic Variation Research and the Clustering of Points in the Plane},
  journal = {Geographical Analysis},
  volume  = {12},
  number  = {3},
  pages   = {205--222},
  year    = {1980},
  doi     = {10.1111/j.1538-4632.1980.tb00031.x}
}

@article{garcia2015,
  author  = {Garcia, Luis P. F. and de Carvalho, Andre C. P. L. F. and Lorena, Ana C.},
  title   = {Effect of Label Noise in the Complexity of Classification Problems},
  journal = {Neurocomputing},
  volume  = {160},
  pages   = {108--119},
  year    = {2015},
  doi     = {10.1016/j.neucom.2014.10.085}
}

@article{ariasgarcia2024,
  author  = {Arias-Garcia, J. and others},
  title   = {Improved Design for Hardware Implementation of Graph-Based Large Margin Classifiers for Embedded Edge Computing},
  journal = {IEEE Transactions on Neural Networks and Learning Systems},
  volume  = {35},
  number  = {1},
  pages   = {1320--1329},
  year    = {2024}
}

@article{hanriottorresbraga2024,
  author  = {Hanriot, V{\'i}tor M. and Torres, Luiz C. B. and Braga, Ant{\^o}nio P.},
  title   = {Multiclass Graph-Based Large Margin Classifiers: Unified Approach for Support Vectors and Neural Networks},
  journal = {IEEE Transactions on Neural Networks and Learning Systems},
  year    = {2024},
  doi     = {10.1109/TNNLS.2024.3420227}
}

@misc{ggclassification,
  author = {{The GGClassification Contributors}},
  title  = {{GGClassification}: {G}abriel Graph Based Large-Margin Classifiers},
  year   = {2020},
  note   = {R package. \url{https://rdrr.io/cran/GGClassification/}}
}

@article{colhak2025,
  author  = {{\c{C}}olhak, F. and others},
  title   = {Accelerating {IoV} Intrusion Detection: Benchmarking {GPU}-Accelerated vs {CPU}-Based {ML} Libraries},
  journal = {arXiv preprint arXiv:2504.01905},
  year    = {2025},
  url     = {https://arxiv.org/abs/2504.01905}
}

@article{zhang2023,
  author  = {Zhang, Y. and others},
  title   = {A {GPU}-Based Computational Framework that Bridges Neuron Simulation and Artificial Intelligence},
  journal = {Nature Communications},
  volume  = {14},
  pages   = {5798},
  year    = {2023}
}

@incollection{eon2021,
  author    = {Eon, W. and others},
  title     = {Deep Learning with {GPU}s},
  booktitle = {Advances in Computers},
  editor    = {Kim, S. and Deka, G. C.},
  volume    = {122},
  pages     = {167--215},
  publisher = {Elsevier},
  year      = {2021}
}

@inproceedings{magni2014,
  author    = {Magni, Alberto and Dubach, Christophe and O'Boyle, Michael},
  title     = {Automatic Optimization of Thread-Coarsening for Graphics Processors},
  booktitle = {Proceedings of the 23rd International Conference on Parallel Architectures and Compilation Techniques (PACT)},
  pages     = {455--466},
  year      = {2014}
}

@inproceedings{nickolls2008,
  author    = {Nickolls, John and Buck, Ian and Garland, Michael and Skadron, Kevin},
  title     = {Scalable Parallel Programming with {CUDA}},
  booktitle = {ACM SIGGRAPH 2008 Classes},
  year      = {2008},
  doi       = {10.1145/1401132.1401152}
}

@article{owens2008,
  author  = {Owens, John D. and Houston, Mike and Luebke, David and Green, Simon and Stone, John E. and Phillips, James C.},
  title   = {{GPU} Computing},
  journal = {Proceedings of the IEEE},
  volume  = {96},
  number  = {5},
  pages   = {879--899},
  year    = {2008}
}

@misc{nvidia2025guide,
  author = {{NVIDIA Corporation}},
  title  = {{CUDA} {C}++ Programming Guide},
  year   = {2025},
  note   = {Version 13.0. \url{https://docs.nvidia.com/cuda/cuda-c-programming-guide/}}
}

@article{jain2010,
  author  = {Jain, Anil K.},
  title   = {Data Clustering: 50 Years Beyond $K$-Means},
  journal = {Pattern Recognition Letters},
  volume  = {31},
  number  = {8},
  pages   = {651--666},
  year    = {2010}
}

@article{ezugwu2022,
  author  = {Ezugwu, Absalom E. and others},
  title   = {A Comprehensive Survey of Clustering Algorithms: State-of-the-Art Machine Learning Applications, Taxonomy, Challenges, and Future Research Prospects},
  journal = {Engineering Applications of Artificial Intelligence},
  volume  = {110},
  pages   = {104743},
  year    = {2022}
}

@article{thorndike1953,
  author  = {Thorndike, Robert L.},
  title   = {Who Belongs in the Family?},
  journal = {Psychometrika},
  volume  = {18},
  number  = {4},
  pages   = {267--276},
  year    = {1953}
}

@book{cormen2002,
  author    = {Cormen, Thomas H. and Leiserson, Charles E. and Rivest, Ronald L. and Stein, Clifford},
  title     = {Introduction to Algorithms},
  edition   = {2nd},
  publisher = {MIT Press},
  year      = {2001}
}

\end{document}